\documentclass[nonacm]{acmart}

\usepackage[utf8]{inputenc}
\usepackage{array,ragged2e}

\usepackage{todonotes}

\newcolumntype{L}[1]{>{\RaggedRight}p{#1}}  
\newcolumntype{R}[1]{>{\RaggedLeft}p{#1}}   
\newcolumntype{C}[1]{>{\Centering}p{#1}}    
\newcolumntype{M}[1]{>{\RaggedRight}m{#1}}  
\newcolumntype{B}[1]{>{\RaggedRight}b{#1}}  

\AtBeginDocument{
  \providecommand\BibTeX{{
    \normalfont B\kern-0.5em{\scshape i\kern-0.25em b}\kern-0.8em\TeX}}}

\acmBooktitle{arXiv}

\begin{document}

\title[Survey about cyberattack protection motivation in higher education]{Survey about cyberattack protection motivation in higher education: Academics at Slovenian universities, 2017}

\author{Luka Jelovčan}
\email{luka.jelovcan@student.um.si}
\affiliation{
  \institution{University of Maribor}
  \country{Slovenia}
}

\author{Simon Vrhovec}
\email{simon.vrhovec@um.si}
\affiliation{
  \institution{University of Maribor}
  \country{Slovenia}
}

\author{Anže Mihelič}
\email{anze.mihelic@um.si}
\affiliation{
  \institution{University of Maribor}
  \country{Slovenia}
}

\begin{abstract}

This paper reports on a study aiming to explore factors associated with motivation of individuals in organizations to protect against cyberattacks. The objectives of this study were to determine how fear of cyberattacks, perceived severity, perceived vulnerability, perceived threats, measure efficacy, self-efficacy, measure costs, mandatoriness and psychological reactance are associated with protection motivation of individuals in organizations. The study employed a cross-sectional research design. A survey was conducted among academics at six Slovenian universities between June and September 2017. A total of 324 respondents completed the survey (7.6 percent response rate) providing for N=255 useful responses after excluding poorly completed responses. The survey questionnaire was developed in English. A Slovenian translation of the survey questionnaire is available.
\end{abstract}

\keywords{PMT, higher education, university, cybersecurity, cyber threat, computer security, internet}

\maketitle

\section{Introduction}

The survey questionnaire was designed to measure theoretical constructs included in the research model. Most items were taken or adapted from existing literature to fit the study's context. Table \ref{table:constructs} presents the theoretical constructs included in the research model, their definition in this research, and sources from which construct items were taken or adapted.

\clearpage

\begin{table}[h!]
\caption{\label{table:constructs}Theoretical constructs in the survey questionnaire.}
\small
\begin{tabular}{L{.24\textwidth}L{.51\textwidth}L{.18\textwidth}}
\toprule
Theoretical construct & Definition in this research & Sources \\
\midrule
Fear of cyberattacks & The level of the individual’s fear of cyberattacks. & \cite{Osman1994} \\
Perceived severity (organization) & The perceived extent of consequences of a successful cyberattack for the organization. & \cite{Liang2010} \\
Perceived vulnerability (organization) & The perceived probability of a successful cyberattack on the organization. & \cite{Liang2010} \\
Perceived severity (individual) & The perceived extent of consequences of a successful cyberattack for the individual. & \cite{Liang2010} \\
Perceived vulnerability (individual) & The perceived probability of a successful cyberattack on the individual. & \cite{Liang2010} \\
Perceived threats & The perceived extent of threats to the individual posed by cyberattacks. & \cite{Liang2010} \\
Measure efficacy & The perceived efficacy of countermeasures against cyberattacks that an individual can implement. & \cite{Chen2016} \\
Self-efficacy & The individual’s self-efficacy when implementing countermeasures against cyberattacks. & \cite{Taylor1995} \\
Measure costs & The perceived costs of implementing countermeasures against cyberattacks. & \cite{Woon2005} \\
Mandatoriness & The perceived mandatoriness of implementing countermeasures against cyberattacks. & \cite{Boss2009} \\
Psychological reactance & The level of the individual's motivation toward the reestablishment of threatened or eliminated personal freedoms. & \cite{Hong1989}\\
Protection motivation & The level of organizational insider’s motivation to implement countermeasures against cyberattacks. & \cite{Jansen2018,Venkatesh2003} \\
\bottomrule
\end{tabular}
\end{table}

\section{Method}

\subsection{Survey instrument}

To test the research model, a survey questionnaire was developed. Adapted and new questionnaire items (i.e., fear of cyberattacks, perceived severity (organization), perceived vulnerability (organization), perceived severity (individual), perceived vulnerability (individual), perceived threats, measure efficacy, self-efficacy, measure costs, mandatoriness, psychological reactance, and protection motivation) were developed by following a predefined protocol.

The questionnaire was first developed in English and then translated into Slovenian by two translators independently. The translators developed the Slovenian questionnaire through consensus. The Slovenian questionnaire has been pre-tested by 2 independent respondents who provided feedback on its clarity. Based on the received feedback, the Slovenian questionnaire was reviewed to remove any ambiguity. Items were reworded, added, and deleted in the pre-test. To ensure the consistency between the Slovenian and English questionnaire, the Slovenian questionnaire was translated back to English. No significant differences in the meaning between the original items in English and back-translations were noticed. The English questionnaire was however reviewed to update the items and to remove any ambiguity based on the back-translation.

Table \ref{table:questionnaire_en} presents the survey questionnaire in English and Table \ref{table:questionnaire_si} presents the Slovenian translation of the survey questionnaire. All items were measured using a 7-point Likert scale as presented in Table \ref{table:likert}. 

\clearpage

\begin{table}[h!]
\caption{\label{table:questionnaire_en}Survey questionnaire items (English original).}
\footnotesize
\begin{tabular}{L{.30\textwidth}L{.65\textwidth}}
\toprule
Construct & Item \\
\midrule
Perceived threats (PT) & PT1. Cyberattacks pose a serious threat to me. \\
 & PT2. Consequences of successful cyberattacks would highly threaten me. \\
 & PT3. Cyberattacks pose a serious danger to me. \\

Perceived severity (organization) (PSo) & PSo1. A successful cyberattack on our organization would greatly jeopardize the privacy of its confidential data. \\
 & PSo2. A lot of our organization's confidential data collected by a successful cyberattack could be misused for criminal purposes. \\
 & PSo3. A lot of our organization's confidential data collected by a successful cyberattack could be misused against it. \\

Fear of cyberattacks (FoC) & FoC1. I am very afraid of cyberattacks. \\
 & FoC2. The prevalence of cyberattacks is terrifying. \\
 & FoC3. Potential losses due to cyberattacks are causing me strong discomfort. \\
 & FoC4. The danger of cyberattacks is alarming. \\

Perceived vulnerability (individual) (PVi) & PVi1. It is very likely that I will be a victim of a cyberattack in the future. \\
 & PVi2. My chances of becoming a victim of a cyberattack are very high. \\
 & PVi3. I strongly feel that I will become a victim of a cyberattack in the future. \\

Perceived severity (individual) (PSi) & PSi1. A successful cyberattack would greatly jeopardize my privacy. \\
 & PSi2. A lot of my personal data collected by a successful cyberattack could be misused for criminal purposes. \\
 & PSi3. A lot of my personal data collected by a successful cyberattack could be misused against me. \\

Perceived vulnerability (organization) (PVo) & PVo1. It is very likely that our organization will become a victim of a cyberattack in the future. \\
 & PVo2. Chances of our organization becoming a victim of a cyberattack are very high. \\
 & PVo3. I strongly feel that our organization will become a victim of a cyberattack in the future. \\

Mandatoriness (M) & M1. I am required to take protective measures against cyberattacks according to the organization’s information security policy. \\
 & M2. It is expected that I play an active role in taking protective measures against cyberattacks. \\
 & M3. It is expected that I strictly adhere to organizational information security policies and procedures for protecting against cyberattacks. \\
 
Psychological reactance (PR) & PR1. When something is prohibited, I usually think: "That's exactly what I am going to do." \\
 & PR2. I consider advice from others as an intrusion. \\
 & PR3. Advice and recommendations induce me to do just the opposite. \\
 & PR4. I resist attempts of others to influence my decisions. \\
 & PR5. Organizational rules regarding the use of protective measures against cyberattacks trigger a sense of resistance in me. \\
 & PR6. I become annoyed when I am not able to make free and independent decisions about protective measures against cyberattacks. \\
 & PR7. I become angry if my freedom of choice regarding protective measures against cyberattacks is restricted. \\
 & PR8. When someone is forcing me to take protective measures against cyberattacks, I feel like doing the opposite. \\

Measure efficacy (ME) & ME1. The success rate of protective measures against cyberattacks is very high. \\
 & ME2. The probability of stopping cyberattacks by taking protective measures is very high. \\
 & ME3. The likelihood of neutralizing cyberattacks by taking protective measures is very high. \\

Self-efficacy (SE) & SE1. I have no problems using protective measures against cyberattacks. \\
 & SE2. Taking a protective measure against cyberattacks is entirely under my control. \\
 & SE3. I have all resources and skills needed to take protective measures against cyberattacks. \\

Measure costs (MC) & MC1. Taking protective measures against cyberattacks is very work-intensive. \\
 & MC2. Taking protective measures against cyberattacks requires a lot of effort. \\
 & MC3. Taking protective measures against cyberattacks is very time-consuming. \\

Protection motivation (PM) & PM1. I intend to take protective measures against cyberattacks in the future. \\
 & PM2. I predict that I will take protective measures against cyberattacks in the future. \\
 & PM3. I plan to take protective measures against cyberattacks in the future. \\
\bottomrule
\end{tabular}
\end{table}

\begin{table}[h!]
\caption{\label{table:questionnaire_si}Survey questionnaire items (Slovenian translation).}
\footnotesize
\begin{tabular}{L{.35\textwidth}L{.60\textwidth}}
\toprule
Construct & Item \\
\midrule
Perceived threats (PT) & PT1. Spletni napadi mi predstavljajo resno grožnjo. \\
 & PT2. Posledice uspešnih spletnih napadov bi me močno ogrozile. \\
 & PT3. Spletni napadi mi predstavljajo resno nevarnost. \\

Perceived severity (organization) (PSo) & PSo1. Uspešen spletni napad bi zelo ogrozil zasebnost zaupnih podatkov naše organizacije. \\
 & PSo2. Ob uspešnem spletnem napadu bi bilo veliko zaupnih podatkov naše organizacije lahko zlorabljenih v kriminalne namene. \\
 & PSo3. Ob uspešnem spletnem napadu bi bilo lahko veliko zaupnih podatkov naše organizacije zlorabljenih zoper našo organizacijo. \\

Fear of cyberattacks (FoC) & FoC1. Zelo se bojim spletnih napadov. \\
 & FoC2. Razširjenost spletnih napadov je zastrašujoča. \\
 & FoC3. Potencialne izgube zaradi spletnih napadov v meni zbujajo močno neprijetnost. \\
 & FoC4. Nevarnost spletnih napadov je zaskrbljujoča. \\

Perceived vulnerability (individual) (PVi) & PVi1. V prihodnosti bom zelo verjetno žrtev spletnega napada. \\
 & PVi2. Obstaja velika verjetnost da postanem žrtev spletnega napada. \\
 & PVi3. Močno se mi zdi, da bom v prihodnosti postal žrtev spletnega napada. \\

Perceived severity (individual) (PSi) & PSi1. Uspešen spletni napad bi zelo ogrozil mojo zasebnost. \\
 & PSi2. Ob uspešnem spletnem napadu bi bilo veliko mojih osebnih podatkov zlorabljenih v kriminalne namene. \\
 & PSi3. Ob uspešnem spletnem napadu bi bilo lahko veliko mojih osebnih podatkov zlorabljenih zoper mene. \\

Perceived vulnerability (organization) (PVo) & PVo1. Naša organizacija bo v prihodnosti zelo verjetno žrtev spletnega napada. \\
 & PVo2. Obstaja velika verjetnost, da naša organizacija postane žrtev spletnega napada. \\
 & PVo3. Močno se mi zdi, da bo naša organizacija v prihodnosti postala žrtev spletnega napada. \\

Mandatoriness (M) & M1. Od mene se zahteva uporaba zaščitnih ukrepov proti spletnim napadom v skladu z informacijsko-varnostno politiko organizacije. \\
 & M2. Od mene se pričakuje igranje aktivne vloge pri uporabi zaščitnih ukrepov proti spletnim napadom. \\
 & M3. Od mene se pričakuje dosledno upoštevanje organizacijske informacijsko-varnostne politike in postopkov za zaščito pred spletnimi napadi. \\
 
Psychological reactance (PR) & PR1. Če je nekaj prepovedano, si navadno mislim: "Točno to bom naredil." \\
 & PR2. Nasvet ostalih dojemam kot vsiljevanje. \\
 & PR3. Nasveti in predlogi v meni sprožijo željo, da naredim ravno nasprotno. \\
 & PR4. Upiram se poskusom, da bi drugi vplivali name in na moje odločitve. \\
 & PR5. Organizacijska pravila o uporabi zaščitnih ukrepov proti spletnim napadom v meni sprožajo občutek odpora. \\
 & PR6. Če ne morem svobodno in neodvisno odločati o zaščitnih ukrepih proti spletnim napadom, postanem nejevoljen. \\
 & PR7. Omejevanje moje svobode odločanja o zaščitnih ukrepih proti spletnim napadom me ujezi. \\
 & PR8. Ko me nekdo sili v uporabo zaščitnih ukrepov proti spletnim napadom, bi najraje naredil ravno obratno. \\

Measure efficacy (ME) & ME1. Zaščitni ukrepi proti spletnim napadom so lahko zelo uspešni. \\
 & ME2. Z uporabo zaščitnih ukrepov proti spletnim napadom je verjetnost zaustavitve spletnih napadov zelo visoka. \\
 & ME3. Verjetnost nevtralizacije spletnih napadov z uporabo zaščitnih ukrepov proti njim je zelo visoka. \\

Self-efficacy (SE) & SE1. Nimam težav pri uporabi zaščitnih ukrepov proti spletnim napadom. \\
 & SE2. Uporabo zaščitnih ukrepov proti spletnim napadom imam povsem pod nadzorom. \\
 & SE3. Imam vsa potrebna sredstva in znanje za uporabo zaščitnih ukrepov proti spletnim napadom. \\

Measure costs (MC) & MC1. Uporaba zaščitnih ukrepov proti spletnim napadom zahteva veliko dela. \\
 & MC2. Uporaba zaščitnih ukrepov proti spletnim napadom zahteva veliko truda. \\
 & MC3. Uporaba zaščitnih ukrepov proti spletnim napadom je zelo zamudna. \\
 
Protection motivation (PM) & PM1. V prihodnosti nameravam uporabljati zaščitne ukrepe proti spletnim napadom. \\
 & PM2. Predvidevam, da bom v prihodnosti uporabljal zaščitne ukrepe proti spletnim napadom. \\
 & PM3. Načrtujem uporabo zaščitnih ukrepov proti spletnim napadom v prihodnosti. \\
\bottomrule
\end{tabular}
\end{table}

\clearpage

\begin{table}[h!]
\caption{\label{table:likert}7-point Likert scale.}
\small
\begin{tabular}{L{.06\textwidth}L{.19\textwidth}L{.19\textwidth}}
\toprule
Score & English & Slovenian \\
\midrule
1 & Strongly disagree & Močno se ne strinjam \\
2 & Disagree & Se ne strinjam \\
3 & Somewhat disagree & Delno se ne strinjam \\
4 & Neutral & Nevtralno \\
5 & Somewhat agree  & Delno se strinjam \\
6 & Agree & Se strnijam \\
7 & Strongly agree & Močno se strinjam \\
\bottomrule
\end{tabular}
\end{table}

\subsection{Data collection}

We conducted the survey with the Slovenian translation of the questionnaire among academics at six Slovenian universities between 13 June 2017 and 3 September 2017. Respondents were recruited through e-mail addresses which were publicly available at the official websites of universities and their departments. The respondents did not receive any compensation for taking the survey. A total of 4,291 e-mails were sent, and 324 respondents completed the survey providing for a response rate of 7.6 percent. After excluding poorly completed responses (responses with over 50 percent of missing values or standard deviation equal to 0 for constructs fear of cyberattacks, perceived severity (organization), perceived vulnerability (organization), perceived severity (individual), perceived vulnerability (individual), perceived threats, measure efficacy, self-efficacy, measure costs, mandatoriness, psychological reactance and protection motivation), we were left with 255 useful responses as presented in Table \ref{table:sample}.

\begin{table}[h!]
\caption{\label{table:sample}Sample with the number of sent invitations, number of responses, and number of useful responses ($N$) after excluding poorly completed responses.}
\small
\begin{tabular}{L{.03\textwidth}L{.11\textwidth}R{.12\textwidth}R{.09\textwidth}R{.04\textwidth}}
\toprule
ID & Name & Sent invitations & Responses & $N$ \\
\midrule
1 & UL AGRFT & 37 & 2 & 1 \\
2 & UL BF & 564 & 31 & 26 \\
3 & UL EF & 127 & 6 & 6 \\
4 & UL FA & 55 & 2 & 2 \\
5 & UL FDV & 165 & 11 & 10 \\
6 & UL FE & 118 & 11 & 5 \\
7 & UL FGG & 199 & 16 & 14 \\
8 & UL FKKT & 180 & 11 & 6 \\
9 & UL FMF & 181 & 11 & 6 \\
10 & UL FPP & 84 & 5 & 4 \\
11 & UL FRI & 121 & 3 & 3 \\
12 & UL FS & 126 & 8 & 6 \\
13 & UL FSD & 26 & 7 & 6 \\
14 & UL FŠP & 91 & 7 & 5 \\
15 & UL NTF & 91 & 9 & 8 \\
16 & UL PEF & 148 & 14 & 13 \\
17 & UL PF & 41 & 2 & 2 \\
18 & UL TEOF & 45 & 5 & 4 \\
19 & UL ZF & 94 & 10 & 10 \\
20 & UM EPF & 69 & 2 & 1 \\
21 & UM FE & 30 & 2 & 2 \\
22 & UM FERI & 276 & 26 & 17 \\
23 & UM FF & 105 & 5 & 3 \\
24 & UM FKBV & 65 & 1 & 1 \\
25 & UM FKKT & 38 & 0 & 0 \\
26 & UM FL & 26 & 2 & 1 \\
27 & UM FNM & 78 & 8 & 6 \\
28 & UM FS & 189 & 26 & 20 \\
29 & UM FT & 33 & 4 & 3 \\
30 & UM FVV & 24 & 7 & 7 \\
31 & UM FZV & 34 & 6 & 3 \\
32 & UM MF & 46 & 1 & 0 \\
33 & UM PEF & 71 & 4 & 4 \\
34 & UM PF & 39 & 0 & 0 \\
35 & UNG FH & 43 & 3 & 3 \\
36 & UNG FN & 13 & 0 & 0 \\
37 & UNG FZO & 35 & 1 & 1 \\
38 & UNG PTF & 18 & 1 & 1 \\
39 & UNG VŠVV & 18 & 1 & 0 \\
40 & UNM FII & 27 & 2 & 2 \\
41 & UNM FIŠ & 31 & 6 & 6 \\
42 & UNM FOŠ & 30 & 6 & 6 \\
43 & UPR FAMNIT & 159 & 8 & 4 \\
44 & UPR FHŠ & 62 & 0 & 0 \\
45 & UPR FM & 64 & 7 & 4 \\
46 & UPR FTŠ & 39 & 6 & 5 \\
47 & UPR PEF & 103 & 13 & 13 \\
48 & VSNM FPUV & 9 & 0 & 0 \\
49 & VSNM FTS & 5 & 0 & 0 \\
50 & VSNM FUPI & 12 & 2 & 2 \\
51 & VSNM FZV & 7 & 3 & 3 \\
\midrule
\multicolumn{2}{l}{Total} & 4,291 & 324 & 255 \\
\bottomrule
\end{tabular}
\end{table}

The first page of the survey is presented in Table \ref{table:first-page}.

\clearpage

\begin{table}[h!]
\caption{\label{table:first-page}The first page of the survey.}
\small
\begin{tabular}{L{.47\textwidth}L{.47\textwidth}}
\toprule
English original & Slovenian translation \\
\midrule
\textit{Self-protection of employees against cyberattacks} \newline ~ \newline Dear Sirs, \newline ~ \newline We invite you to participate in a study on self-protection against cyberattacks which is carried out by the Faculty of Security Sciences at the University of Maribor. This study aims to gain an insight into the factors that influence the decisions of employees to take measures to protect against malicious software. Estimated time to complete the survey is 4-5 minutes. \newline ~ \newline For additional information regarding the study, contact us at simon.vrhovec@fvv.uni-mb.si. & \textit{Samovarovanje zaposlenih pred spletnimi napadi} \newline ~ \newline Spoštovani, \newline ~ \newline  vabimo vas k sodelovanju v raziskavi o samovarovanju pred spletnimi napadi, ki jo izvajamo na Fakulteti za varnostne vede Univerze v Mariboru. Z raziskavo želimo pridobiti vpogled v dejavnike, ki vplivajo na odločitve zaposlenih o sprejemanju ukrepov za zaščito pred škodljivo programsko opremo. Predviden čas izpolnjevanja ankete je 4-5 minut.
 \newline ~ \newline Za dodatne informacije v zvezi z raziskavo nam pišite na simon.vrhovec@fvv.uni-mb.si. \\
\bottomrule
\end{tabular}
\end{table}

\section{Results}

\subsection{Sample}

Demographic characteristics of the sample are presented in Table \ref{table:demographics}.

\begin{table}[h!]
\caption{\label{table:demographics}Demographic characteristics of the sample.}
\small
\begin{tabular}{L{.18\textwidth}L{.24\textwidth}R{.08\textwidth}}
\toprule
Characteristic & Value & Frequency \\
\midrule
Gender & 1 -- Male & 106 \\
 & 2 -- Female & 134 \\
 & \textit{Missing} & 84 \\

Age group & 21-25 & 8 \\
 & 26-30 & 26 \\
 & 31-35 & 36 \\
 & 36-40 & 29 \\
 & 41-45 & 42 \\
 & 46-50 & 24 \\
 & 51-55 & 30 \\
 & 56-60 & 16 \\
 & 61-65 & 20 \\
 & 66-70 & 5 \\
 & 71 or more & 5 \\
 & \textit{Missing} & 83 \\

Formal education & High school or less & 4 \\
 & Bachelor's degree & 7 \\
 & Master's degree & 64 \\
 & PhD degree & 163 \\
 & \textit{Missing} & 86 \\
\bottomrule
\end{tabular}
\end{table}

\clearpage

\subsection{Frequencies}

Frequencies of all variables for measured theoretical constructs are presented in Table \ref{table:Frequencies}.

\begin{table}[h!]
\caption{\label{table:Frequencies}Frequencies of variables.}
\small
\begin{tabular}{L{.07\textwidth}R{.03\textwidth}R{.03\textwidth}R{.03\textwidth}R{.03\textwidth}R{.03\textwidth}R{.03\textwidth}R{.03\textwidth}R{.07\textwidth}R{.07\textwidth}R{.07\textwidth}}
\toprule
Variable & 1 & 2 & 3 & 4 & 5 & 6 & 7 & Valid & Missing & Total \\
\midrule
PT1 & 8 & 31 & 19 & 26 & 81 & 106 & 48 & 319 & 5 & 324 \\
PT2 & 3 & 30 & 18 & 26 & 71 & 102 & 67 & 317 & 7 & 324 \\
PT3 & 7 & 35 & 19 & 24 & 83 & 102 & 47 & 317 & 7 & 324 \\
PSo1 & 6 & 10 & 14 & 25 & 51 & 94 & 109 & 309 & 15 & 324 \\
PSo2 & 15 & 53 & 24 & 38 & 70 & 60 & 42 & 302 & 22 & 324 \\
PSo3 & 13 & 44 & 25 & 34 & 84 & 61 & 41 & 302 & 22 & 324 \\
FoC1 & 13 & 56 & 30 & 46 & 81 & 48 & 19 & 293 & 31 & 324 \\
FoC2 & 9 & 30 & 26 & 45 & 54 & 83 & 43 & 290 & 34 & 324 \\
FoC3 & 12 & 51 & 30 & 43 & 67 & 58 & 30 & 291 & 33 & 324 \\
FoC4 & 6 & 18 & 19 & 26 & 73 & 100 & 48 & 290 & 34 & 324 \\
PVi1 & 23 & 52 & 33 & 62 & 51 & 31 & 3 & 255 & 69 & 324 \\
PVi2 & 19 & 60 & 33 & 57 & 47 & 40 & 5 & 261 & 63 & 324 \\
PVi3 & 23 & 73 & 31 & 63 & 42 & 23 & 5 & 260 & 64 & 324 \\
PSi1 & 10 & 45 & 33 & 34 & 73 & 56 & 21 & 272 & 52 & 324 \\
PSi2 & 21 & 65 & 35 & 45 & 56 & 35 & 10 & 267 & 57 & 324 \\
PSi3 & 21 & 57 & 33 & 43 & 68 & 37 & 10 & 269 & 55 & 324 \\
PVo1 & 8 & 34 & 26 & 61 & 55 & 39 & 12 & 235 & 89 & 324 \\
PVo2 & 6 & 34 & 29 & 58 & 46 & 42 & 19 & 234 & 90 & 324 \\
PVo3 & 8 & 44 & 32 & 57 & 52 & 31 & 11 & 235 & 89 & 324 \\
M1 & 7 & 23 & 11 & 33 & 42 & 81 & 51 & 248 & 76 & 324 \\
M2 & 10 & 28 & 20 & 37 & 51 & 69 & 32 & 247 & 77 & 324 \\
M3 & 5 & 17 & 14 & 30 & 51 & 67 & 65 & 249 & 75 & 324 \\
PR1 & 114 & 99 & 4 & 14 & 7 & 4 & 3 & 245 & 79 & 324 \\
PR2 & 66 & 135 & 16 & 13 & 10 & 1 & 2 & 243 & 81 & 324 \\
PR3 & 91 & 122 & 12 & 10 & 6 & 1 & 2 & 244 & 80 & 324 \\
PR4 & 32 & 73 & 36 & 47 & 19 & 26 & 11 & 244 & 80 & 324 \\
PR5 & 61 & 116 & 17 & 30 & 10 & 5 & 4 & 243 & 81 & 324 \\
PR6 & 56 & 114 & 19 & 22 & 17 & 10 & 4 & 242 & 82 & 324 \\
PR7 & 62 & 117 & 15 & 22 & 12 & 12 & 3 & 243 & 81 & 324 \\
PR8 & 100 & 110 & 8 & 14 & 4 & 3 & 2 & 241 & 83 & 324 \\
ME1 & 2 & 4 & 4 & 13 & 50 & 120 & 36 & 229 & 95 & 324 \\
ME2 & 1 & 6 & 6 & 17 & 61 & 102 & 35 & 228 & 96 & 324 \\
ME3 & 2 & 7 & 8 & 21 & 61 & 101 & 27 & 227 & 97 & 324 \\
SE1 & 3 & 4 & 11 & 21 & 28 & 111 & 53 & 231 & 93 & 324 \\
SE2 & 7 & 20 & 17 & 53 & 72 & 50 & 10 & 229 & 95 & 324 \\
SE3 & 15 & 45 & 31 & 36 & 60 & 40 & 7 & 234 & 90 & 324 \\
MC1 & 5 & 34 & 23 & 40 & 58 & 45 & 11 & 216 & 108 & 324 \\
MC2 & 4 & 34 & 27 & 39 & 55 & 45 & 11 & 215 & 109 & 324 \\
MC3 & 9 & 39 & 30 & 48 & 48 & 36 & 7 & 217 & 107 & 324 \\
PM1 & 3 & 3 & 1 & 25 & 44 & 110 & 40 & 226 & 98 & 324 \\
PM2 & 2 & 3 & 2 & 21 & 36 & 118 & 47 & 229 & 95 & 324 \\
PM3 & 7 & 7 & 12 & 45 & 31 & 91 & 31 & 224 & 100 & 324 \\
\bottomrule
\end{tabular}
\end{table}

\section{Discussion}

This paper presents the results of a survey about cyberattack protection motivation in higher education. Future studies may focus on other factors associated with protection motivation of individuals in organizations. Such studies would be beneficial both to better explain protection motivation of organizational insiders, and to better understand the associations between different factors associated with it.

\section*{Acknowledgements}

We would like to express our sincere gratitude to the respondents who took their time to participate in our survey.

\bibliographystyle{ACM-Reference-Format}
\bibliography{main}

\end{document}